\documentclass[12pt, letterpaper]{article}

\pdfoutput=1

\usepackage{amssymb, graphics, epsfig}
\usepackage{epsf}
\usepackage{epstopdf}
\usepackage{color}
\usepackage{amsmath,amsthm,amsfonts}
\usepackage{graphicx}
\usepackage{physics}
\usepackage{amssymb}

\usepackage{hyperref}
\hypersetup{
	colorlinks   = true, 
	urlcolor     = blue, 
	linkcolor    = magenta, 
	citecolor    = blue   
}
\graphicspath{{figs/}}

\newcommand{\nc}{\newcommand}
\nc{\ba}{\begin{eqnarray}}
\nc{\ea}{\end{eqnarray}}
\newcommand\be{\begin{equation}}
\newcommand\ee{\end{equation}}

\newcommand{\bea}{\begin{eqnarray}}
\newcommand{\eea}{\end{eqnarray}}

\def\bfk{{\bf k}}

\def\bfq{{\bf q}}

\def\Ylm{Y_{\ell m}(\theta, \phi) }

\def\zlm{Z_{\ell m}}

\begin{document}

\vspace{5mm}
\vspace{0.5cm}
\begin{center}

\def\thefootnote{\fnsymbol{footnote}}

{\bf     White Hole Cosmology and Hawking Radiation from\\ Quantum  Cosmological Perturbations}
\\[0.5cm]

{  Hassan Firouzjahi$\footnote{firouz@ipm.ir}$, Alireza Talebian$\footnote{talebian@ipm.ir}$ }   \\[0.5cm]
{\small \textit{ School of Astronomy, Institute for Research in Fundamental Sciences (IPM) \\ P.~O.~Box 19395-5531, Tehran, Iran}}

\vspace{0.1cm}

\end{center}

\vspace{.8cm}

\hrule \vspace{0.3cm}


\begin{abstract}
The spacetime inside the white hole is like an anisotropic cosmological background with the  past singularity  playing the role of a big bang singularity. The scale factor along the extended spatial direction is contracting while the scale factor along the two-sphere is expanding. We consider an eternal Schwarzschild manifold and study quantum cosmological perturbations generated near  the white hole singularity which  propagate  towards the past event horizon and to exterior of the black hole.  It is shown that an observer deep inside the white hole and an observer far outside the black hole both share the same vacuum. We calculate the Hawking radiation associated to these quantum white hole perturbations as measured by an observer in the exterior of the black hole. Furthermore, we  also consider the Hawking radiation for the general case where the initial cosmological perturbations deep inside the white hole are in ``non-vacuum" state yielding to  a deviation from Planck distribution. This analysis suggests that if the black hole is not entirely black (due to Hawking radiation) then the white hole is not entirely white either. \\

\end{abstract}
\vspace{0.5cm} \hrule
\def\thefootnote{\arabic{footnote}}
\setcounter{footnote}{0}
\newpage
\section{Introduction}

Black hole (BH) has played important roles in the developments of theoretical physics. Furthermore, during the past decades  numerous observations  have suggested the existence of massive and supermassive black holes at the center of typical galaxies.  The recent detections of gravitational waves by the LIGO and VIRGO collaborations  \cite{LIGOScientific:2016aoc, LIGOScientific:2016sjg, LIGOScientific:2018jsj}  from the merging of binary astrophysical black holes put the reality of black holes in cosmos beyond doubt.  On the theoretical side BH physics play key roles in understanding of quantum gravity which were studied extensively  in the past decades.

The physical processes governing the dynamics of the  interior of  the BH are not well understood. The prime reason is that  the interior  region of BH is causally disconnected from the exterior region. Any in-falling signal smoothly passes through the event horizon and no signal can escape from inside the event horizon to an outside observer.  
This phenomena suggests that  the interior of a BH is like a cosmological background bounded by the event horizon.   This interpretation is supported from the fact that inside the horizon, the roles of coordinates $ t$ and $r$ as the time-like and space-like coordinates are switched. There have been works in the past to treat the interior of a BH as a cosmological background. For example 
the idea that the  interior of BH may be replaced by a non-singular dS space-time was studied in   \cite{Sakharov:1966aja, Poisson:1988wc, Frolov:1989pf, Frolov:1988vj, Firouzjahi:2016nle, wenda1988junction, Brandenberger:2021ken, Gaztanaga:2021rnv, Gaztanaga:2022gbd}, see also  \cite{Sato:1981bf, Maeda:1981gw, Sato:1981gv, Farhi:1986ty, Blau:1986cw, Oshita:2016btk}
for similar ideas but in  somewhat different contexts. The main motivation  to replace  the interior of BH by the dS background was to remove  the singularity of BH. On the physical ground one may expect that the singularity of BH is a shortcoming of the classical  general relativity. On very small scales, say on Planck scale, it is expected that the quantum gravity effects can not be neglected. 
It is expected that these effects provide mechanisms to resolve the singularity inside the BH. 
For example, the idea of maximum curvature of space-time  
\cite{markov1982limiting, Frolov:1989pf, Frolov:1988vj} is an interesting proposal in this direction.

Like the interior of the BH, the white hole (WH) background is more akin to a cosmological spacetime in which  the global structure of spacetime suggests that the past singularity $r=0$ behaves as the onset of big bang singularity in which  the signal generated from  $r=0$ inside the WH 
propagates towards the future  null infinity ${\cal I}^{+}$. The WH spacetime can be viewed as an anisotropic cosmological background with its spatial part having 
the topology $R\times S^2$ known as the Kantowski-Sachs \cite{Kantowski:1966te} spacetime.  On the other hand, perturbations in FLRW  cosmological backgrounds have been studied extensively. Indeed, it is believed that all structures in observable Universe are generated from tiny quantum fluctuations generated during primordial inflation. It is therefore a natural question to study perturbations  inside the WH as a particular cosmological background.   With these discussions in mind, in this work first we study some basic cosmological properties of the WH geometry as part of an eternal 
Schwarzschild manifold. 
Then we study the quantum perturbations of a test scalar field  generated deep inside the WH which will propagate towards the past event horizon and eventually reaching to an observer far  outside the BH. We calculate the spectrum of  Hawking radiation  
as measured by this observer.  For a related study see also \cite{Chakraborty:2015nwa}.

It is believed that the WHs are not stable and have disappeared in early universe 
\cite{Eardley:1974zz} so the current analysis may not be directly relevant to observable Universe. However, we treat WH as part of an eternal BH manifold which exists along with BH as required by the  time reversal symmetry of general relativity. We find interesting properties of WH cosmology as an anisotropic cosmological background 
while studying quantum field theory in WH background can provide  a non-trivial example of  quantum field theory in curved backgrounds. 

Hawking radiation in the space-time of WH has been also studied in \cite{Jusufi:2017nib,Giavoni:2020gui}. In Ref. \cite{Jusufi:2017nib}, it was shown that there is a Hawking radiation associated to a WH spacetime equal to the BH Hawking temperature when viewed from the outside region of the WH geometry based on BH-to-WH tunneling scenario~\cite{Barcelo:2016hgb} during the gravitational collapse process. The tunnelling rate inside the horizon of a WH for scalar and massive vector particles were calculated using the Hamilton-Jacobi method. Our study differs form  \cite{Jusufi:2017nib} in two ways: first we deal with an eternal WH while in Ref.~\cite{Jusufi:2017nib} the WH is considered as a long-lived remnant of BH during gravitational collapse. Second, we study the quantum fluctuations generated near the WH singularity which propagate naturally towards the past event horizon. While in \cite{Jusufi:2017nib}, Hawking radiation viewed as a quantum tunnelling effect to the tunnelling rate of particles. Moreover, the authors of \cite{Giavoni:2020gui} have presented a generalization of the Hawking effect for dynamical trapping horizons by calculating the tunnelling rate in the Hamilton-Jacobi formalism. They studied the quantum effects (quantum tunnelling) across various horizons in general, including the WH horizon (past outer trapping horizon).

\section{Background Geometry}
\label{prelim}

In this section we briefly review the necessary preliminaries from BH background  and set the stage for the WH cosmology.

\subsection{Black hole preliminaries}
\label{prelim}

We consider the Schwarzschild metric with the following line element 
\ba
\label{metric-Sch}
\dd s^2 = - \big(
1- \frac{2GM}{r}\big) \dd t^2 + \frac{\dd r^2}{\big(
	1- \frac{2GM}{r}\big)} + r^2 \dd \Omega^2 \, ,
\ea
in which $G$ is the Newton constant, $M$ is the mass of the BH as measured by an observer at infinity while $\dd\Omega^2$ represents the metric on a unit two-sphere. The coordinate system $(r,t)$ only covers the exterior of the whole manifold while becoming singular on the BH event horizon at $r= r_S\equiv 2 GM$. Naively speaking, for the interior of the BH and the WH the roles of $t$ and $r$ coordinates are reversed in which $t$ becomes spacelike while $r$ becomes timelike. This also suggests that the interiors of BH and WH are actually dynamical, mimicking cosmological backgrounds. 

To cover the entire manifold, we can use the  Kruskal coordinate in which 
\ba
\label{metric-Kruskal}
\dd s^2 =  \frac{32 G^3M^3}{r} e^{ {-r}/{2GM} } \big( -\dd T^2 + \dd R^2 \big) +   r^2 \dd \Omega^2 \, ,
\ea
where $(T, R)$ coordinate is related to the original  coordinate $(t, r)$ by
\ba
\label{UV-r}
T^2 - R^2 =  e^{r/2GM} \Big(
1- \frac{r}{2GM}
\Big)  \, ,
\ea
and
\ba
\label{UV-t}
\frac{T}{R}  = \tanh(\frac{t}{4 G M}) \, .
\ea
It is evident that $T$ is timelike while $R$ is spacelike throughout. 

It is also very convenient to use the  Kruskal coordinate in its lightcone base
$(U, V)$ defined via 
\ba
U\equiv GM( T-R),  \quad  \quad V\equiv GM( T+R) \, ,
\ea
in which the line element takes the following form 
\ba
\label{metric-Kruskal}
\dd s^2 = -\frac{32 G M}{r} e^{-r/2GM} \dd U \dd V +   r^2 \dd \Omega^2 \, .
\ea
Note that in our convention, the coordinates $(U, V)$ carry the dimension of length (or time). 

\begin{figure}[t]
\begin{center}
	\includegraphics[scale=0.3]{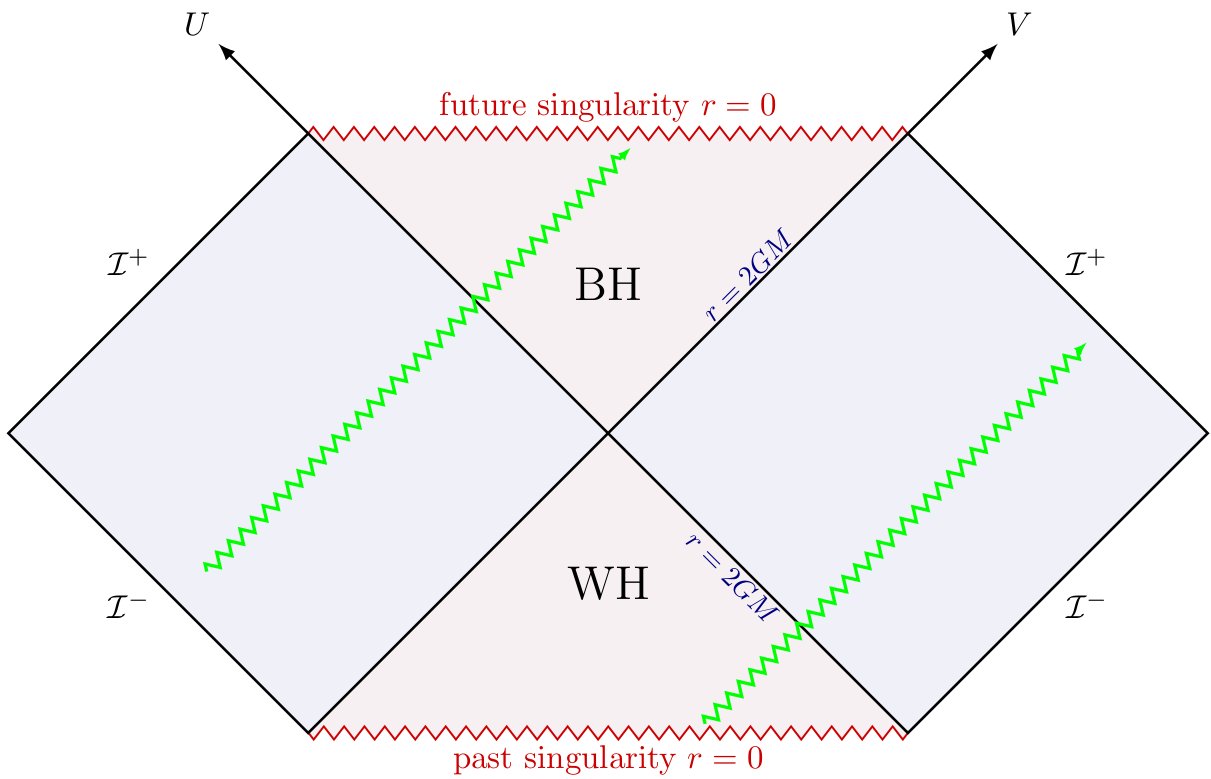}
	\end{center}
\caption{ The Kruskal diagram. We are interested in quantum perturbations generated from the past singularity $r=0$ inside the WH, crossing the past horizon at $V=0, U<0$ 
and propagating towards ${\cal I}^{+}$. To have a complete Cauchy surface of initial conditions, we also consider the perturbations generated from left ${\cal I}^{-}$ and right ${\cal I}^{-}$ located respectively to the left and to the right of WH.
The wavy green line represents the propagation of the right mover modes $e^{\pm i \omega u}$  while the left movers $e^{\pm i \omega v}$  are not shown for brevity. 
}
\label{Kruskal}
\end{figure}

A conformal diagram of the entire Schwarzschild manifold is presented in Fig. \ref{Kruskal}.
The exterior of the BH is the region $V>0, U <0$ while the interior of the BH is the region $U, V>0$ bounded by the  future singularity 
$r=0$. The WH region is given by $U, V<0$ bounded by the past singularity $r=0$. Both the interior of the BH and the WH share the common property that their backgrounds are dynamical corresponding to anisotropic cosmological setups. However, the crucial difference is that the singularity of the BH is in future, i.e. it represents a big crunch singularity while the singularity of the WH is in the past, i.e. it represents a big bang singularity. As mentioned before, the WH is unstable and may not exist in current observable Universe. However, in our treatment we consider an eternal BH in which 
a WH is an integral part of the full manifold. In this eternal background, the WH exists along with BH as required by the time reversal symmetry of the classical general relativity.  

Defining the tortoise coordinate $\dd r_* = (1- \frac{2GM}{r})^{-1} \dd r$ for the interior of BH and the WH regions ($r<2GM$) we have
\ba
\label{r*-WH}
r_* = r + 2 GM \ln \left( 1- \frac{r}{2GM} \right) ,   \qquad \quad ({r < 2 GM}) \, .
\ea
Restricting our attention to WH region, we note that  $-\infty < r_* \leq 0$ in which $r_*=-\infty$ corresponds to the past event horizon $r=2GM, V=0$ with $U<0$
while $r_* =0$ corresponds to the past singularity at $r=0$.

\subsection{White hole cosmology}
\label{WH-cosmology}

As mentioned above, the WH background represents an anisotropic cosmological setup known as the Kantowski-Sachs background 
with the structure of 
$R \times S^2$.   To see this more clearly, let us define the future directed time coordinate $\dd\tau=-\dd r_*$ so
\ba
\label{tau-def}
\tau = -r - 2 GM \ln \left( 1- \frac{r}{2GM} \right) \, .
\ea
Correspondingly, the WH singularity is at $\tau=0$ while the past event horizon  $U<0, V=0 $ is mapped to $\tau=+\infty$. Now defining the spacelike coordinate 
$\dd x \equiv \dd t$ with $-\infty < x < +\infty$, 
the metric in the WH region takes the following cosmological form
\ba
\label{WH-metric}
\dd s^2 = a(\tau)^2 ( -\dd \tau^2 + \dd x^2) + r(\tau)^2 \dd \Omega^2 \, ,
\ea
in which the scale factor $a(\tau)$ is defined via
\ba
\label{a-scale}
a(\tau) \equiv  \bigg( \frac{2 G M}{r(\tau)} -1  \bigg)^{\frac{1}{2}} \, . 
\ea
It is important to note that $r= r(\tau)$.

The metric (\ref{WH-metric}) represents an anisotropic background
with two scale factors $a(\tau)$ and $r (\tau)$. At the ``big bang"
singularity $\tau=r=0$,  we have $a(\tau) \rightarrow \infty$ while $r(\tau) \rightarrow 0$ so the space along the $x$ direction starts off very large and contracts as time pass by while the scale factor along the two-sphere 
is originally zero and stars to expand. 

It is insightful to look at the metric in two limiting time $\tau \rightarrow 0$ and 
$\tau \rightarrow \infty$ more closely. Using Eq. (\ref{tau-def}), at earlier time when 
$r \rightarrow 0$  we have $r^2 \simeq 4 G M \tau$ so the metric takes the following form
\ba
\label{limit1}
\dd s^2 \simeq  \sqrt{\frac{GM}{\tau}} ( - \dd\tau^2 + \dd x^2) + 4 G M \tau ~\dd \Omega^2
\quad \quad (\tau \rightarrow 0) \, .
\ea
The contraction along the $x$ direction and the expansion along the two-sphere is evident from the above line element.

On the other hand, near the past horizon $r=2GM, V=0 $ we have 
\ba
r(\tau) \simeq 2 GM \Big[ 1- \exp \big(-\frac{\tau}{2 GM} \big)
\Big] \quad \quad (\tau \rightarrow \infty) \, ,
\ea
so the line element takes the following form
\ba
\label{limit2}
\dd s^2 \simeq \exp \big({\frac{-\tau}{2 G M}} \big) ( - \dd\tau^2 + \dd x^2) + 4 G^2 M^2  \dd \Omega^2
\quad \quad (\tau \rightarrow \infty) \, .
\ea
We see that near the horizon, the two-sphere reaches the fixed radius $2 GM$
while the scale factor in the $x$-direction falls off exponentially.  Although the scale factor $a(\tau)$ falls off exponentially, but the spacetime is regular near the horizon. Indeed, calculating the Riemann tensor associated to the metric (\ref{limit2}) 
we find that $R_{\theta \phi \theta \phi} = 4 G^2 M^2 \sin(\theta)^2$ while all other components are zero so the background is regular near the horizon. 
Of course, this was expected with the understanding that 
the apparent singularity of the metric (\ref{metric-Sch}) near the horizon was a mere coordinate singularity.  

While the coordinate $\tau$ is the conformal time, we can define the ``cosmic time"
$\hat t$ related to the conformal time via $\dd \hat t \equiv a(\tau) ~\dd \tau$ so 
\ba
\label{hat t-r-eq}
\dd \hat t = \frac{\dd r}{ \sqrt{\frac{2 G M}{r} -1}} \, .
\ea
Choosing the onset of big bang singularity to be at $\hat t=0$, we obtain
\ba
\hat t= - \sqrt{r ( 2 G M -r)} + M G  \Big[\frac{\pi}{2}-  \mathrm{arc} \sin \Big( 1- \frac{r}{G M} \Big)  \Big] \, ,
\ea
so $ 0<    \hat t < {\pi} GM$ inside the WH.  
From the inverse of the above expression, we can express $r= r(\hat t)$ as well so the metric in term of cosmic time is given by
\ba
\label{metric-cosmic}
\dd s^2 = - \dd \hat t^2 + \Big( \frac{2 G M}{r (\hat t\, )} - 1 \Big) \dd x^2 + 
r (\hat t\, )^2 \dd \Omega^2. 
\ea

It is also  instructive to look at the geometry near the singularity when given in the form of Eq. (\ref{metric-cosmic}).  Near $r=0$, we have 
\ba
r \simeq ( G M)^\frac{1}{3}  \bigg(\frac{3 \hat t}{\sqrt2} \bigg)^\frac{2}{3}
\quad \quad (\hat t \rightarrow 0) \, ,
\ea
so the metric (\ref{metric-cosmic}) becomes
\ba
\label{metric-cosmic2}
\dd s^2 \simeq -  \dd \hat t^2 +  ( G M)^\frac{2}{3} \Big( c_1 {\hat t}^{-\frac{2}{3}} \dd x^2 + c_2 {\hat t}^{\frac{4}{3}} \dd \Omega^2 \Big) \, ,
\ea
where $c_1$ and $c_2$ are some numerical factors. One can check that the above geometry has the form of the Kasner  metric. More specifically, the Kasner metric is an exact solution of the Einstein field equation in the vacuum 
with the following form (in four dimension), 
\ba
\label{Kasner}
\dd s^2 = - {\dd \hat t\, }^2 + \sum_{i=1}^3 {\hat t \,}^{2 p_i} \dd x_i^2 \,  ,
\ea
in which the exponents $p_i$ satisfy the following constraints:
\ba
\label{constraints-Kasner}
\sum_{i=1}^3 p_i = \sum_{i=1}^3 p_i^2 =1 \, .
\ea
The structure of the above two constraints suggest that at least one of the 
exponents $p_i$ should be negative.  In our example of WH cosmology near the singularity with the metric (\ref{metric-cosmic2}) we obtain $p_1= -\frac{1}{3}$ while 
$p_2= p_3 =\frac{2}{3}$. One can check that these exponents indeed satisfy the constraints (\ref{constraints-Kasner}) with the contraction being along the 
{$x$} direction.

We can also define the Hubble ``expansion" rate associated to metric (\ref{metric-cosmic}).  Actually, we  have two Hubble  rates as we have two scale factors $a(\hat t)$ and $r(\hat t)$. Defining the Hubble rate associated to the scale factor $a(\hat t)$ by $H_1\equiv \frac{1}{a}  \frac{\dd a}{ \dd \hat t}$, we obtain
\ba
H_1= -\frac{G M}{{r (\hat t )}^2} \bigg( \frac{2 G M}{r(\hat t\, )} -1 \bigg)^{-\frac{1}{2}} \, .
\ea 
Since the scale factor $a$ is contracting, the Hubble  rate $H_1$ is negative. 
In addition, $H_1$ diverges both near the WH singularity and near the horizon. The former divergence is genuine as the spacetime is singular at $r=0$ so physical quantities like energy density or Hubble rate diverge.   However, the 
divergence  of $H_1$ near the past event horizon is an artifact of coordinate singularity as can be seen from the relation between $\dd \hat t$ and $\dd r$ in Eq. (\ref{hat t-r-eq}).

On the other hand, defining the Hubble rate associated to the scale factor $r(\hat t)$ via 
$H_2\equiv \frac{1}{r(\hat t) }  \frac{\dd r(\hat t) }{ \dd \hat t}$, we obtain
\ba
H_2= \frac{1}{r(\hat t)} \bigg( \frac{2 G M}{r(\hat t \, )} -1 \bigg)^{\frac{1}{2}} \, .
\ea
One can check that the following relation between the two Hubble rates hold,
\ba
H_1 H_2 = -\frac{G M}{r (\hat t\, )^3} \, .
\ea
We see that $H_2$ diverges near the singularity as expected but it vanishes near the past event horizon. This is because near the horizon ($\tau \rightarrow \infty$) the radius of the  two-sphere approaches a  constant size so $H_2$ vanishes in this limit.


\section{Quantum Field in White Hole Geometry }
Here we study the quantum field perturbations  in a WH background. We primarily consider a real massless scalar field $\Phi$ while the final results will be easily extended to fields of other spins such as the vector and tensor perturbations.  
We work in the test field limit where the backreaction of the field on the background is negligible. We use the 
metric (\ref{WH-metric}) with the conformal time $\tau \equiv -r_*$ while $x\equiv t$ represents the extended spatial direction.   

As the background enjoys the two-dimensional rotation invariance we  expand the quantum field $\Phi$ in terms of the spherical harmonics  $Y_{\ell m} (\theta, \phi)$ as follows 
\ba
\label{Phi-Z}
\Phi = \frac{1}{r} \sum_{\ell =0}^\infty \sum_{ m=-\ell}^\ell   Z_{\ell m}(\tau, x) Y_{\ell m} (\theta, \phi)  \, ,
\ea
in which $Z_{\ell m}(\tau, x)$ play the role of the quantum mode operator. In addition, 
 we have pulled out a factor $1/r$ such that $Z_{\ell m}$ will be
the canonically normalized field in the perturbation analysis.  Note that the reality of $\Phi$ requires that $Z_{\ell m}^* = (-1)^m Z_{\ell\,  -m}$. 

Using the orthogonality of the spherical harmonics, 
\ba
\int \dd \Omega\,  \Ylm Y_{\ell' m'}^*(\theta, \phi) = \delta_{\ell \ell'} \delta_{m m'} \, , 
\ea
and the relations, 
\ba
\label{sum-Ylm}
Y_{l m}(\theta, \phi)^* = (-1)^m  Y_{l \,  -m}(\theta, \phi) \, ,
\quad \quad 
\sum_{m=-\ell}^{\ell} | Y_{\ell m} (\theta,\phi)|^2 =\frac{2 \ell +1}{4 \pi} \, ,
\ea
the action of the scalar field takes the following canonical form
\ba
\label{action-canonic}
S=\frac{1}{2} \int \dd \tau \dd x  \sum_\ell
\Bigg[  \big | \frac{\partial Z_{\ell m}}{\partial \tau} \big|^2
- \big | \frac{\partial Z_{\ell m}}{\partial x} \big|^2 + 
\Big(1- \frac{2G M}{r(\tau)} \Big) \Big( \frac{2 G M}{r(\tau)^3} + \frac{\ell (\ell +1) }{r(\tau)^2}\Big)
|Z_{\ell m}|^2
\Bigg] \, .
\ea
From the above action one obtains the scalar field perturbation equation, 
\ba
\label{eq-tort-1}
\partial_{\tau}^2  \zlm - \partial_x^2 \zlm  -   \Big(1- \frac{2 G M}{r(\tau)} \Big)  \left[ \frac{2 G M}{r(\tau)^3} + \frac{\ell (\ell +1) }{r(\tau)^2}
\right]  \zlm =0  \, .
\ea
The above equation is similar to the  standard Regee-Wheeler perturbation equation for the field in the exterior region of the BH. However, in our setup where WH is treated as a cosmological background, the above equation may be interpreted as the extension of the  Sasaki-Mukhanov equation to an anisotropic cosmological background such as the Bianchi I universe.

Since the spatial coordinate $x$ has the  translation invariance we can Fourier expand  $\zlm \propto e^{-i \bfk x}$ in which $\bfk$ is a one dimensional Fourier mode. Plugging the above mode expansion in our master equation (\ref{eq-tort-1}),  the equations of the perturbations take the form of  the Regge-Wheeler equation~\cite{Regge:1957td}, 
\ba
\label{eq-pert}
\partial_{\tau}^2  \zlm + \big( k^2 - V_{\mathrm{eff}}(\tau) \big) \zlm = 0 \, ,
\ea 
with the following effective potential
\ba
\label{Veff}
V_{\mathrm{eff}} (\tau) = \Big(1- \frac{2 G M}{r(\tau)} \Big)  \left[ \frac{\ell (\ell+1)}{r(\tau)^2} + \frac{2 G M}{r(\tau)^3} \right] \, .
\ea 

{It is worth mentioning that for the axial metric perturbations 
and the electromagnetic	field perturbations the effective potential in 
Eq. \eqref{eq-pert}  takes the following forms, 
\ba
\label{Veff_spin}
V_{\mathrm{eff}} (\tau) = \Big(1- \frac{2 G M}{r(\tau)} \Big)  \left[ \frac{\ell (\ell+1)}{r(\tau)^2} + \frac{2 G M}{r(\tau)^3}(1-s^2) \right] \, ,
\ea 
in which $s$ is the spin of the corresponding field, i.e. $s=0,1,2$  for the
scalar, electromagnetic and gravitational fields respectively.}

We see that $V_{\mathrm{eff}} (\tau)<0$ such that it diverges near $\tau \rightarrow 0$ while it goes to zero  near the horizon corresponding to $\tau \rightarrow \infty$. With the relation between $\tau$ and $r$ given in Eq. (\ref{tau-def}) it is not possible to solve the above mode function. However, near the singularity the contribution of the term containing $\ell( \ell+1)$ can be neglected compared to the last term in $V_{\mathrm{eff}} (\tau)$ which is more divergent. Also, near the horizon, the effective potential vanishes so the contributions of the whole potential is subleading compared to the  constant term $k^2$. As a result, we can neglect the contribution of the 
term containing $\ell( \ell+1)$ in the mode function equation (\ref{eq-pert}) with reasonable accuracies. Correspondingly, to simplify the notation, we drop the subscripts $\ell, m$ in the following analysis and effectively work in the $s$-wave limit with  $\ell=m=0$.

We treat the wave equation as quantum fluctuations generated at the point of past singularity $r=0$ (i.e. Big Bang) inside the WH. To emphasis, note that the WH region is bounded by the past horizon $V=0, U<0$, the future horizon $U=0, V<0$ and the past singularity $r=0$, see Fig.  \ref{Kruskal}. After the generation of quantum perturbation  at past singularity $r=0$, they propagate towards the past horizon $V=0, U<0$ and eventually towards ${\cal I}^+$ which is measured by an observer at future infinity in the exterior part of the BH.  
However, to have a complete Cauchy initial condition, we also have to consider the perturbations generated from left ${\cal I}^-$ and right ${\cal I}^-$ (which are  respectively to the left and to the right of WH),  see Fig.  \ref{Kruskal} for a schematic view.  We would like to see if the perturbations generated deep inside the WH region is related to Hawking radiation \cite{Hawking:1975vcx, Unruh:1976db} as measured by  the remote observer in future infinity  at ${\cal I}^{+}$.  For a review of Hawking radiation see \cite{Unruh:1983ms, Mukhanov:2007zz, Townsend:1997ku, fabbri2005modeling, Visser:2001kq, Jacobson:2003vx}.

Expanding the mode function as  $Z \sim e^{i \bfk x} g_k( \tau)$, and neglecting the contribution of $\ell(\ell+1) $ as discussed above, 
the wave equation for the mode $g_k(\tau)$  is given by
\ba
\label{tor-mode-eq0}
g_k''(\tau)  + \left[  k^2 -   \frac{2 G M (1- s^2)}{r(\tau)^3} \Big(1- \frac{2GM}{r(\tau)} \Big) \right]  g_k(\tau)=0  \, ,
\ea
in which a prime on $g_k(\tau)$ denotes the derivative with respect to $\tau$.

As mentioned before, this equation is in par with the equation for cosmological perturbations in standard FLRW cosmology. So interesting insights can be obtained 
when comparing to perturbations in FLRW background.  As reviewed in previous section, we have two scale factors $r(\tau)$ and $a(\tau)$, in which the latter is contracting. Since we decomposed the perturbations into spherical harmonics and restricted to $s$-wave limit, then Eq. (\ref{tor-mode-eq0}) represents the dynamics of perturbations in a 1+1  contracting universe with the  scale factor $a(\tau)$ given in Eq. (\ref{a-scale}). Like in the studies of perturbations in FLRW cosmology, we
can divide the perturbations into long ``superhorizon" and short ``subhorizon" perturbations  depending on the value of the physical wave vector $k/a$. The structure of  Eq. (\ref{tor-mode-eq0}) suggests that  for a given $r(\tau)$ the perturbations which satisfy $\frac{k^2}{a^2}< \frac{G M}{r(\tau)^3}$ are outside the horizon. Curiously, we see that at the  start of singularity near $r=0$, all perturbations were superhorizon. As time goes by and the scale factor $a(\tau)$ contracts (i.e. $r(\tau)$ increases)  the perturbations enter the  horizon and  $\frac{k^2}{a^2}> \frac{G M}{r(\tau)^3}$. This is the hallmark of a contracting universe in contrast to an inflationary background where the perturbations of cosmological interests are initially deep inside the horizon and then stretched outside the horizon by the exponential expansion of the background. An interesting conclusion from Eq. (\ref{tor-mode-eq0}) is that perturbations with the physical wavelength smaller than $(\frac{2 GM}{r_S^3})^{-\frac{1}{2}}= r_S$ remain subhorizon throughout the whole evolution.

With the above qualitative discussions we try to solve  Eq. (\ref{tor-mode-eq0}) approximately 
as it can not be solved analytically for the whole  region $0 < \tau <\infty$. 
We note that for the regions near the singularity $r=0$, we have $ r^2  \simeq 4 GM \tau$ and Eq.  \eqref{tor-mode-eq0} simplifies to
\ba
\label{tor-mode-eq1}
g_k''(\tau)  + \left(  k^2 + \frac{1-s^2}{4 \tau^2}  \right)  g_k(\tau)=0 \, 
\quad \quad (\tau \rightarrow 0) \, .
\ea
The above equation can be solved analytically, yielding 
\ba
\label{gk-sol}
g_k (\tau) = \sqrt{\tau} \left[ C_1(\omega) H_{\frac{s}{2}}^{(1)}(\omega \tau) + C_2(\omega)  H_{\frac{s}{2}}^{(2)}( \omega \tau)  
\right] \, , \quad \quad (\tau \rightarrow 0) \, ,
\ea
in which $H_{s/2}^{(1, 2)}$ are the Hankel functions, $C_1(\omega), C_2(\omega)$ are constants of integrations and $\omega \equiv  |k|\ge   0$.

The solution \eqref{gk-sol}  can not be used for the whole region  $0< \tau< \infty$. However, as we move away from $r=0$, the effective potential in wave equation (\ref{tor-mode-eq0}) falls off rapidly and the dominant term in the big bracket in  Eq. (\ref{tor-mode-eq0}) is simply $k^2$. This suggests that the solution  (\ref{gk-sol}) is a very good solution for both  regions   $ \tau \rightarrow 0$ and $\tau \rightarrow \infty$ with modifications in the intermediate region. Therefore, we take  the analytic solution (\ref{gk-sol}) to be qualitatively valid for the entire region $0 < \tau <\infty$. 

To fix the constants $C_1(\omega), C_2(\omega)$ we look at the initial conditions of perturbations.  In our view, these are quantum fluctuations generated near the
past singularity $r=0$  deep inside the WH. This interpretation is quite similar to the case of FLRW cosmology, in which the observed large scale perturbations are generated from quantum fluctuations in early universe at the time of big bang (actually deep inside a Hubble patch during inflation). Indeed, the link between these two views is quite  natural since the geometry inside the WH 
is dynamical, representing an anisotropic cosmological background, as reviewed in subsection \ref{WH-cosmology}.

With these discussions in mind, we expand the quantum mode as follows
\ba
\label{mode-Z-tor}
Z (\tau, x) = \int_{-\infty}^{+\infty} \frac{\dd \bfk}{\sqrt{2 \pi}} 
\left[  e^{i \bfk x} g_k(\tau) b_{\bfk} +   e^{-i \bfk x} g^*_k(\tau) b_{\bfk}^\dagger  \right] \, ,
\ea
in which $b_\bfk$ and $b_{\bfk}^\dagger$  are the annihilation and the creation operators defined deep in the WH background (i.e. near $r=0$)
with the usual commutation relations
\ba
[ b_\bfk, b^\dagger_{\bfk'}] = \delta (\bfk - \bfk') \, , \quad 
[ b_\bfk, b_{\bfk}] = [ b^\dagger_\bfk, b^\dagger_{\bfk'}] = 0 \, .
\ea
Note that inside the WH, the spacelike coordinate $x$ spans the range $(-\infty, +\infty)$ so the wave number $\bfk$ 
in the integral (\ref{mode-Z-tor}) varies in the interval   $(-\infty, +\infty)$.

To find the proper normalization, we have to impose the equal time commutation relation between the field $Z$ and its conjugate momentum $\partial_\tau Z$, 
\ba
\label{quan.}
[ Z (\tau, x) , \partial_{\tau} Z(\tau, x') ] =  i \delta (x-x') \, .
\ea
 Imposing the above quantization condition on the mode function (\ref{gk-sol}) yields
\ba
\label{norm}
\big| C_2(\omega) \big|^2 - \big|C_1(\omega) \big|^2 = \frac{\pi}{4} \, .
\ea
The above constraint does not uniquely fix $C_1(\omega)$ and $C_2(\omega)$. However, the above relation  suggests the natural choice $C_1=0$ and $C_2 = \sqrt \pi/2$.
This choice may be compared to the standard ``Bunch-Davies" vacuum in a dS background 
which carries the lowest energy among all possible vacua. Of course one may consider the general case in which both $C_1(\omega)$ and $C_2(\omega)$ are present, parallel to  the ``non Bunch-Davies" vacuum in inflation.  To start, first we consider the natural case $C_1=0$ for simplicity. In Section \ref{C12-general}, we extend the analysis for general values of $C_1(\omega)$ and $C_2(\omega)$. 
Now, choosing $C_1=0$, we have
\ba
\label{g*-rt}
g_k (\tau) = \sqrt{\frac{\pi \tau}{4}}  H_{s/2}^{(2)}( \omega \tau)  \, .
\ea

The mode propagates towards the exterior of WH into the past event horizon $\tau \rightarrow \infty$ with $U<0, V=0$. 
Using the asymptotic expansion of $H_{{s/2}}^{(2)}( \omega \tau)$, near the past horizon we obtain
\ba
\label{Hankel-asym}
H_{{s/2}}^{(2)}( \omega \tau) \rightarrow \sqrt{\frac{2}{ \pi {\omega} \tau}} e^{i \big(-\omega \tau + \frac{\pi}{4}{(1+s)} \big) } \quad \quad (\tau \rightarrow \infty) \, .
\ea
Plugging the above asymptotic expansion in the mode expansion (\ref{mode-Z-tor}) we obtain
\ba
\label{Z-approx}
Z(\tau, x)\!\rightarrow\!\int_0^\infty\! \frac{\dd \omega }{\sqrt{4 \pi \omega}}  
{\Big[
e^{ i \omega (-\tau + x )+i \frac{\pi}{4} } b_\omega +  e^{- i \omega (-\tau + x ) -i \frac{\pi}{4}} b_\omega^\dagger 
+ e^{ i \omega (-\tau - x )+i \frac{\pi}{4} } b_{-\omega} +  e^{- i \omega (-\tau - x ) -i \frac{\pi}{4}}b_{-\omega}^\dagger 
\Big]
}
\ea

The above mode function is the superposition of two independent left and right 
moving free waves, depending respectively on the combinations $\tau+x$ and 
$\tau-x$. Defining the tortoise lightcone {coordinates}, 
\ba
\label{uv-WH}
u\equiv \tau + x  , \quad \quad  v\equiv  \tau -x \, ,
\ea
and rescaling  $b_\bfk$ and $b_\bfk^\dagger$ with the unimportant phase factor,  $b_\bfk \rightarrow e^{\frac{i\pi}{4} } b_\bfk$ and  
$b_\bfk^\dagger \rightarrow e^{-\frac{i\pi}{4} } b_\bfk^\dagger $,  we have 
\ba
\label{wave-rt1}
Z(\tau, x)  {|_{r \rightarrow 2GM }} = \int_0^\infty \frac{\dd \omega }{\sqrt{2 \pi}} \frac{1}{\sqrt{2 \omega }} \left[
e^{-i \omega v }  b_\omega  +  e^{+i \omega v }  b_\omega^\dagger 
+ e^{-i \omega u }  b_{-\omega}  +  e^{+i \omega u }  b_{-\omega}^\dagger 
\right] \, .
\ea
It is reassuring that our analytic solution (\ref{mode-Z-tor}) nicely coincides with the notion of the right movers $e^{\pm i \omega u}$  and  left movers $e^{\pm i \omega v}$  expansion of the mode function near the horizon.  The right mover corresponds to wave crossing the past event horizon $V=0, U<0$,  propagating towards the right ${\cal I}^+$ in BH region. On the other hand,  the left mover corresponds to wave crossing  the future event horizon $U=0, V<0$, propagating towards the left ${\cal I}^+$ on the mirror side of the exterior of BH. 

Our goal is  to examine what an observer outside the BH on the right $\cal I^+$ with $V \rightarrow +\infty$  measures. For this purpose, we need to choose the right mover from the superposition in Eq. (\ref{wave-rt1}).  Considering the right mover part of Eq. (\ref{wave-rt1}) near the past horizon we have  
\ba
\label{wave-rt}
Z(u)  {|_{r \rightarrow 2 GM}}\rightarrow \int_0^\infty \frac{\dd \omega }{\sqrt{2 \pi}} \frac{1}{\sqrt{2 \omega }} \left[ 
 e^{-i \omega u }  b_{-\omega}  +  e^{+i \omega u }  b_{-\omega}^\dagger 
\right] \, .
\ea
{This solution, generated in the  WH region, can be used as the initial condition to propagate from the surface $V=0$ to outside of the BH on  the surface $\cal I^+$ where the measurements are  made.} After crossing the past horizon, 
this wave will be scattered by the effective potential in  the exterior region $2GM < r < \infty$. However, the crucial difference compared to the WH region is that the effective Regge-Wheeler  potential  for the exterior of BH  is not divergent and only has a finite bump. Therefore, at least  for large enough value of $\omega$,  one can ignore the effects of the Regge-Wheeler  potential and treat the solution emerging from the WH  as simply propagating without change towards $\cal I^+$. Because of this reason, we practically consider the near horizon limit of the solution for the outgoing solution in the exterior region. 

Now the crucial point is that the tortoise coordinate $r_*$ and equivalently the time coordinate $\tau$ covers only the WH region. In order to properly extend the solution (\ref{wave-rt}) through the past horizon and into the exterior region, we have to pass to the Kruskal coordinate which smoothly covers {\it both} the WH and the BH regions. In other words, we employ the wave function solution in the Kruskal coordinate to smoothly connect the wave function for the exterior region  $r > 2 GM$ to the solution in the WH region as given by Eq. (\ref{wave-rt}).  The Kruskal coordinate has the unique property that  it is suitable 
for a freely falling observer near the horizon while the  mode expansion (\ref{wave-rt}) mixes the positive and the negative frequencies of the freely falling observer  which is the main reason behind the Hawking radiation. 

Near the horizon the coordinate $(U, V)$ is best suitable for a freely falling observer in which the geometry is nearly approximated by
\ba
\dd s^2 \simeq -16G M \dd U \dd V + (2 GM)^2 \dd \Omega^2  \quad \quad ( r \rightarrow 2 GM) \, .
\ea
Ignoring the angular parts, it describes a two-dimension flat spacetime. The wave equation in $(U, V)$ coordinate in near horizon approximation is simply  
\ba
\label{UV-eq1}
\partial_U \partial_V Z(U, V)  =0 \,  , \quad \quad ( r \rightarrow 2 GM) \, .
\ea
This admits the expected plane wave solution $e^{\pm i \omega U}$ and $e^{\pm i \omega V}$. Considering the right mover solution for our purpose 
we have 
\ba
\label{wave-U}
Z(U)= \int_0 ^\infty \frac{\dd \omega}{\sqrt{2 \pi} } \frac{1}{\sqrt {2 \omega}} \left( e^{- i \omega U} a_\omega +
e^{ i \omega U} a_\omega^\dagger
\right) \, , \quad \quad (r \rightarrow 2 G M), 
\ea
in which now the operators $a_\omega$ and $a^\dagger_\omega$ are the annihilation and the creation operators defined for a freely falling observer near the horizon equipped with the Kruskal coordinate.  The important point is that the solution (\ref{wave-U}) is valid on both sides of past event horizon, whether $V>0$ or $V<0$. 
Therefore, we can expand the solution (\ref{wave-rt})
in the base of  the solution (\ref{wave-U}). This allows us to find $b_\omega $ in Eq. (\ref{wave-rt}) in the WH region  in terms of $a_\omega$ and $a_\omega^\dagger $ in Eq. (\ref{wave-U}). Finally, we can use Eq. (\ref{wave-U}) as the initial condition for the exterior region to be used as the initial conditions for the observer on right $\cal I^+$. 

To connect the two solutions (\ref{wave-U}) and (\ref{wave-rt}) we have to 
find a dictionary between $(u, v)$ and $(U, V)$ coordinates.  In the WH region both $U$ and $V$ are negative and they are related to $(u, v)$ coordinate via
\ba
\label{UV-uv1}
V= - \frac{1}{4\kappa} e^{-\kappa v} \, , \quad U= -\frac{1}{4\kappa} e^{-\kappa u} \, ,
\ea
in which $\kappa \equiv \frac{1}{4 GM}$ is the surface gravity.

\section{  Hawking Radiation}
\label{UV-coordinate}

For a freely falling observer near the horizon the natural coordinate to employ is the Kruskal coordinate which is non-singular. On the other hand, an observer 
using the coordinate $(\tau, x)$ is accelerating with respect to the Kruskal coordinate. Correspondingly, the vacua employed by these two observers are not identical. Specifically, a positive frequency mode function in (\ref{wave-rt}) mixes the negative and positive frequency modes of the Kruskal vacuum. The interpretation is that,  the Kruskal vacuum, as viewed by an accelerator observer in WH, is not empty and carries particles with Planck distribution. In this section we look at this issue more closely both from the point of view of an observer 
deep in WH region and also from the point of view of the far observer in right  $\cal I^+$.

\subsection{Relation between $ (b_{-\tilde\omega}, b^\dagger_{-\tilde\omega}) $ and $(a_\omega, a^\dagger_\omega)$ }
\label{WH-Bogol}

Our first job is to find a relation between $(b_{-\tilde\omega}, b^\dagger_{-\tilde\omega})$ and $(a_\omega, a^\dagger_\omega)$. Here to simplify the notation, we denote the frequency in the mode function for the 
WH region by $\tilde \omega$. Also see Appendix \ref{inner-prod} for an alternative derivation of $(b_{-\tilde\omega}, b^\dagger_{-\tilde\omega})$ based on the standard inner product approach. 

From Eq. (\ref{wave-rt}) we have
\ba
\label{b-Omega}
b_{-\tilde \omega} = \sqrt{2 \tilde\omega} \int_{-\infty}^{+\infty} \frac{\dd u}{\sqrt{2 \pi}}\,  e^{i \tilde \omega u} Z \, .
\ea
Now,  expressing $Z$ in terms of the $U$ coordinates as given in Eq. (\ref{wave-U}), this leads to
\ba
\label{b-omega2}
b_{-\tilde \omega} = \int_0^\infty \dd \omega \Big( \alpha_{\omega \tilde \omega}
a_\omega + \beta_{\omega \tilde\omega} a_\omega^\dagger
\Big) \, ,
\ea
in which the Bogoliubov coefficients $\alpha_{\omega \tilde \omega}$ and
$\beta_{\omega \tilde \omega}$ are defined via 
\ba
\label{alpha-def}
\alpha_{\omega \tilde \omega} \equiv \sqrt{\frac{\tilde \omega}{\omega}} 
\int_{-\infty}^{+\infty} \frac{\dd u}{2 \pi} 
e^{i \tilde \omega u} e^{- i \omega U} \, ,
\ea
and
\ba
\label{beta-def}
\beta_{\omega \tilde \omega} \equiv \sqrt{\frac{\tilde \omega}{\omega}} 
\int_{-\infty}^{+\infty} \frac{\dd u}{2 \pi} 
e^{i \tilde \omega u} e^{ i \omega U} \, .
\ea

Imposing the standard commutation relations among the two sets of operators
$(a_\omega, a^\dagger_\omega)$ and $(b_{\tilde\omega}, b^\dagger_{\tilde\omega})$ one can check that the Bogoliubov coefficients satisfy the following normalization condition, 
\ba
\label{normalization1}
\int_0^\infty \dd \omega \Big( \alpha_{\omega \tilde \omega}  \alpha_{\omega \tilde \omega'}^* -  \beta_{\omega \tilde \omega}  \beta_{\omega \tilde \omega'}^*
\Big) = \delta( \tilde \omega - \tilde \omega^\prime) \, .
\ea
In particular, for the special case $\tilde \omega= \tilde \omega'$,  the above normalization condition yields
\ba
\label{normalization2}
\int_0^\infty  \dd\omega \Big( |\alpha_{\omega \tilde \omega}|^2 -  
|\beta_{\omega \tilde \omega}|^2  
\Big) = \delta( \bf{0} ) \, .
\ea
Note that the factor $\delta ({\bf 0})$ appears because we have used the infinite length normalization along the spatial $x$ direction.   In a real situation 
it would be replaced by the physical length.

\begin{figure}[t]
\begin{center}
	\includegraphics[scale=0.85]{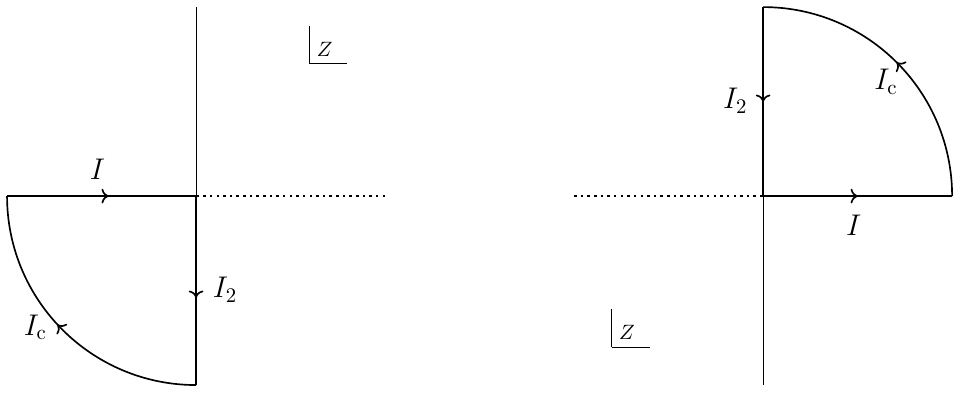}
	\end{center}
\caption{ Left: the contour in the complex plane to calculate the {integral $I$} given in Eq.  (\ref{int-b}) for $-\infty < U <0$.  The integral  $I_c$ evaluated along the part of large circle $C$ vanishes  while the integral $I_2$ over the negative imaginary axis yields to $\Gamma$ function given in Eq. (\ref{Gamma}). The branch cut  associated with the function {$(-U)^{-1-i \frac{\tilde \omega}{\kappa}}$ is on the positive real axis} denoted by the dotted line. Right: the contour to calculate the integral Eq. (\ref{int-c}) with the same description as in left figure, but now $0< U < \infty$ where 
the branch cut is on negative real axis and the contour is closed on upper complex plane.  }
\label{contour}
\end{figure}

To calculate the integrals appearing in $\alpha_{\omega \tilde \omega}$ and
$\beta_{\omega \tilde \omega}$ we use the relation {$U= - \frac{1}{4\kappa} 
	e^{-\kappa u}$} yielding $du= -\frac{1}{\kappa} dU/U$ in which $U$ ranges in the interval $(-\infty, 0)$.   This gives, 
\ba
\label{int-b}
I\equiv  \int_{-\infty}^{+\infty} \dd u ~e^{i \tilde \omega u} e^{- i \omega U}  &=&   \frac{1}{\kappa}\int_{-\infty}^0 \dd (4 \kappa U) (- 4 \kappa U)^{-1- \frac{i\tilde \omega}{\kappa}} e^{-i \omega U} \nonumber\\
&=&
 \frac{1}{\kappa} e^{ \frac{\pi \tilde \omega}{2\kappa} } \Big( \frac{\omega}{4\kappa} \Big)^{ \frac{i\tilde \omega}{\kappa}} \Gamma \big( \frac{-i\tilde \omega}{\kappa} \big) \, .
\ea
The contour in the complex plane to calculate the above integral is shown in left panel of Fig. \ref{contour} where the branch cut  associated with the function 
$(-U)^{-1-i \frac{\tilde \omega}{\kappa}}$ is on the positive real axis. The integral over the part of large circle $C$
with negative imaginary value vanishes while the integral on the imaginary axis 
denoted by $I_2$ yields the $\Gamma$ function given by
\ba
\label{Gamma}
\Gamma(z) = \int_0^{+\infty} \dd x ~ e^{-x} x^{z-1} \, .
\ea
Similarly, for the other integral in $\beta_{\omega \tilde \omega}$ we have  
\ba
 \int_{-\infty}^{+\infty} \dd u ~ e^{i \tilde \omega u} e^{ i \omega U}  = 
  \frac{1}{\kappa} e^{ -\frac{\pi \tilde \omega}{2\kappa} } \Big( \frac{\omega}{4\kappa} \Big)^{ \frac{i\tilde \omega}{\kappa}} \Gamma \big( \frac{-i\tilde \omega}{\kappa} \big) \, ,
\ea
in which now the contour $C$  of the large circle  closes on the upper complex planes with positive imaginary values. 

Combining the above results yield 
\ba
\label{alpha-beta-WH}
 \left \{ \begin{array}{cc} 
 \alpha_{\omega \tilde\omega} \\ 
 \beta_{\omega \tilde\omega}
\end{array} \right.
= \frac{1}{2 \pi \kappa}  \sqrt{\frac{\tilde \omega}{\omega}}  e^{ \pm \frac{\pi \tilde \omega}{2\kappa} } \,  \Big( \frac{\omega}{4\kappa} \Big)^{ \frac{i\tilde \omega}{\kappa}} \Gamma \big( \frac{-i\tilde \omega}{\kappa} \big) \, .
\ea
From the above expressions we find that 
\ba
\label{alpha-beta}
| \alpha_{\omega \tilde\omega} |^2 =  e^{ \frac{2\pi \tilde \omega}{\kappa} }  
| \beta_{\omega \tilde\omega} |^2 \, .
\ea
Plugging this relation into the normalization condition (\ref{normalization2}) yields the following identity 
\ba
\label{beta-identity}
\int_0^\infty \dd \omega ~ | \beta_{\omega \tilde\omega} |^2 
= \frac{\delta ({\bf 0})}{ e^{ \frac{2\pi \tilde \omega}{\kappa} } -  1 } \, ,
\ea
which will be useful later on. 

Having calculated $b_{-\tilde \omega }$, we can obtain 
$b_{-\tilde \omega }^\dagger$ from the complex conjugation of
Eq. (\ref{b-omega2}).

\subsection{Mode function outside the white hole}
\label{outside-WH}

Our next job is to  propagate the mode function from the WH region 
across the past horizon to the  exterior region and then to $\cal I^+$.
As the Kruskal coordinate is continuous across the horizon, we can simply use 
the mode function (\ref{wave-U}) for the observer outside the WH region.

For the exterior of WH, the spacetime is static so the $t$ coordinate is timelike while $r$ is spacelike as usual. The tortoise coordinate is now given 
by $r_* = r+ 2 G M \ln(\frac{r}{2 GM}-1)$, in which the past horizon  is mapped 
to $r_*=-\infty$ while for the far observer $r_*=+\infty$.  Furthermore, the tortoise light-cone coordinates $(u, v)$ are given by $u\equiv t-r_*$ and $v\equiv t+r_*$ such that 
the Kruskal light-cone coordinates $(U, V)$ are now given by 
\ba
\label{UV-uv2}
V=  \frac{1}{4\kappa} e^{\kappa v} \, , \quad U= -\frac{1}{4\kappa} e^{-\kappa u} \, .
\ea
As we are dealing with the right mover solution with the mode function $Z= Z(U)$, we need to express the  right mover solution across the past horizon from the WH region to the exterior region.  Looking at the quantum mode solution of the Regee-Wheeler equation  for the observer 
near the past horizon in the exterior region one can see that the right mover solution has the following form 
\ba
\label{wave-u-BH}
Z(u)= \int_0 ^\infty \frac{\dd \tilde{\omega}}{\sqrt{2 \pi} } \frac{1}{\sqrt {2 \tilde \omega}} \left( e^{- i \tilde \omega u} d_{\tilde \omega} +
e^{ i \tilde \omega u} d_{\tilde\omega}^\dagger
\right) \, , \quad \quad (r \rightarrow 2 G M^+)
\ea
in which the creation and annihilation operators $d_{\tilde \omega}$ and $d_{\tilde\omega}^\dagger$ are defined for this observer just outside the WH. 

Our goal is to express the operators $d_{\tilde \omega}$ and $d_{\tilde\omega}^\dagger$ in terms of the corresponding operators in WH region, i.e. $b_{-\tilde \omega}$ and $b_{-\tilde\omega}^\dagger$. Following a very similar step as in the  previous subsection, we can express $d_{\tilde \omega}$ and $d_{\tilde\omega}^\dagger$ in terms of the Kruskal operators 
$a_{ \omega}$ and $a_{\omega}^\dagger$. However, now the crucial point is that the relation between the $U$ and $u$ coordinates as given by
Eqs. (\ref{UV-uv2}) and (\ref{UV-uv1}) are identical. Therefore, the Bogoliubov coefficients relating $d_{\tilde \omega}$ and $d_{\tilde\omega}^\dagger$ to 
$a_{ \omega}$ and $a_{\omega}^\dagger$ are the same which relates $b_{-\tilde \omega}$ and $b_{-\tilde\omega}^\dagger$ to  
$a_{ \omega}$ and $a_{\omega}^\dagger$. Therefore, we conclude that 
\ba
\label{b-d}
d_{\tilde \omega} = b_{-\tilde \omega} \, , \quad \quad
d_{\tilde \omega}^\dagger = b_{-\tilde \omega}^\dagger \, .
\ea
Interestingly, we see that the observers on both sides of the past horizon agree on the form of the mode function and share the same vacuum. This may be expected since locally the horizon surface is regular so any mode function from the WH region can continuously cross the horizon to the exterior region without modification.

In the limit where we neglect the bump of the potential in the Regee-Wheeler equation the quantum mode function propagates feely from the near horizon region to the observer in far region. Correspondingly, the observer in 
$\cal I^+$ detects the mode function (\ref{wave-u-BH}). In addition, because of the relations (\ref{b-d}),  the observer  in  $\cal I^+$ shares the same vacuum as the observer deep  inside the WH.  While these two observers  share the same vacuum, but their vacuum is different than the vacuum defined by a locally inertial observer using the Kruskal coordinate. This is because the Bogoliubov 
coefficients $\alpha_{\omega \tilde \omega}$ and $\beta_{\omega \tilde \omega}$ are non-trivial. As such, both of these two observers conclude that the vacuum defined by the Kruskal coordinate is not empty.

Let us denote the vacuum defined by the  Kruskal observer  by $|0\rangle_{\rm K}$ which is annihilated by $a_\bfk$, $a_\bfk |0\rangle_{\rm K} =0$. Also let us denote  
 the vacuum defined deep inside the WH  by $|0\rangle_{\rm WH}$ which is 
 annihilated by  $b_\bfq $,   $b_\bfq |0\rangle_{\rm WH} =0$. As shown above, this vacuum is also annihilated by $d_\bfq$ employed by the observer in the BH region, i.e. $d_\bfq |0\rangle_{\rm WH} =0$.  Note that  $|0\rangle_{\rm WH}$ was used to define the positive frequency solution (\ref{g*-rt}) yielding to (\ref{wave-rt}).  As mentioned before, the two states $|0\rangle_{\rm K}$ and  $ |0\rangle_{\rm WH}$ are different.  For the observer which defines  $|0\rangle_{\rm WH}$ as the vacuum inside the WH, the state  $|0\rangle_{\rm K} $ is not empty. This is the source of Hawking radiation.   Now, we  calculate the number of produced particles $_{\rm K}\langle 0| b_{-\tilde \omega}^\dagger b_{-\tilde \omega} | 0 \rangle_{K} = {_{K}}\langle 0| d_{\tilde \omega}^\dagger d_{\tilde \omega} | 0 \rangle_{K}$ as detected by either the WH observer or the observer at $\cal I^+$.

Using the  formulae (\ref{b-omega2}) for $b_{-\tilde \omega}$ and similarly for $b^\dagger_{-\tilde \omega}$, we obtain
\ba
_{\rm K}\langle 0| b_{-\tilde \omega}^\dagger b_{-\tilde \omega} | 0 \rangle_{K} &=&
\int_0^\infty \dd \omega \dd \omega' ~ \beta_{\omega \tilde \omega}^* \beta_{\omega' \tilde\omega}\,   {_{\rm K}\langle} 0| a_{ \omega} a_{ \omega}^\dagger | 0 \rangle_{K} \nonumber\\
&=& \int_0^\infty \dd \omega ~| \beta_{\omega \tilde \omega} |^2 \, .
\ea
Now, using the relation (\ref{beta-identity}), we conclude
\ba
_{\rm K}\langle 0| b_{-\tilde \omega}^\dagger b_{-\tilde \omega} | 0 \rangle_{K} 
= \frac{\delta ({\bf 0})}{ e^{ \frac{2\pi \tilde \omega}{\kappa} } -   1 } \, .
\ea

Calculating the number density  measured by  the observer in BH region (say at at $\cal I^+$), we obtain 
\ba
\frac{_{\rm K}\langle 0| d_{ \tilde \omega}^\dagger d_{ \tilde \omega} | 0 \rangle_{K}}{ \delta ({\bf 0})} =
\frac{_{\rm K}\langle 0| b_{-\tilde \omega}^\dagger b_{-\tilde \omega} | 0 \rangle_{K}}{ \delta ({\bf 0})} = \frac{1}{e^{ \frac{2\pi \tilde \omega}{\kappa} } -1} \, ,
\ea
which is the expected Planck spectrum with the Hawking temperature,
\ba
T_H = \frac{\kappa}{2 \pi} = \frac{1}{8 \pi G M}. 
\ea
Consequently, the observer outside the WH  measures a flux of particle with the 
Planck distribution emitted from WH. Alternatively, the above equation also suggests that an observer deep in WH region  absorbs particles with the same Planckian distribution from the exterior region.  Naively speaking, if the BH is not completely black (via Hawking radiation) then the WH is not completely white either! This conclusion is consistent with the results of 
\cite{Giavoni:2020gui} who have studied  the Hawking effect for dynamical trapping horizons,  calculating the quantum tunnelling rate using the Hamilton-Jacobi formalism. 

Note that we worked in the simplified picture that the effects of the Regee-Wheeler potential and the scattering of the propagating waves towards $\cal I^+$ are neglected. This is a good approximation  for large enough $\omega$  in which one may ignore the effect of the scattering by the effective potential. In general,  the flux  measured by the remote observer is essentially a Planck spectrum but with a greybody factor correction.

\subsection{Outgoing wave from left ${\cal I}^-$}

As mentioned before, to have a complete Cauchy initial conditions, we also have to consider perturbations generated from
left ${\cal I}^-$ and right ${\cal I}^-$ which respectively are to the left and to the right of exterior of  WH, see Fig. \ref{Kruskal}.  The perturbations generated from left ${\cal I}^-$  are right movers, i.e. the mode functions depend on $u$ or $U$ coordinates,  while the perturbations generated at right ${\cal I}^-$  are left movers, depending on $v$ or $V$ coordinates. Since from the start we have considered the right movers, 
here we consider the right moving perturbations originated from the left ${\cal I}^-$. The discussions for the left movers will be the same.

The region enclosed by the left ${\cal I}^-$ and the part of future horizon with $V=0, U>0$ (which is usually referred to as  part of the ``other universe") is  similar to the exterior of BH in our part of the universe. The only difference is that from the surface ${\cal I}^-$ to past event horizon the future directed time is $-t$. To see this, let us construct the Kruskal coordinate in this region. Knowing that $V<0$ and $U>0$, the Kruskal coordinate is given by
\ba
\label{Kruskal2}
V= - \frac{1}{4\kappa} e^{-\kappa v} \, , \quad U= \frac{1}{4\kappa} e^{\kappa u} \, ,
\ea
in which 
\ba
\label{uv-new}
v= -(t + r_*) \quad , \quad u= -t + r_* \, ,
\ea
and now 
\ba
r_* \equiv r+ 2 GM \ln \left(\frac{r}{2GM}-1  \right). 
\ea
Note that $r_*$ defined above has a branch cut with respect to $r_*$ defined inside the WH given in Eq. (\ref{r*-WH}). 
As $u$ and $v$ change in the range $(-\infty, +\infty)$, $U$ and $V$ respectively change in the range $(0, +\infty)$ and 
$(-\infty, 0)$. On the surface ${\cal I}^-$,  $V\rightarrow -\infty$ so from Eq. (\ref{UV-t}) we find $t\rightarrow +\infty$ while near past event horizon where $V\rightarrow 0$, $t\rightarrow -\infty$. This suggests that the future directed time in this region (outside and to the left of WH) is indeed $-t$. 

In this region, like the exterior of BH, $-\infty < r_* < +\infty$ in which $r_* \rightarrow +\infty$ is near ${\cal I}^-$ while 
$r_* \rightarrow -\infty$ is mapped to the past horizon. As in BH case, the effective potential is given by the Regge-Wheeler 
potential which has a finite bump. Therefore, for high enough frequencies, we can ignore the effects of the potential and the wave equation in $(r_*, t)$ coordinate for the entire region is approximately given by
\ba
\label{eq-tort2}
\partial_{r_*}^2  Z - \partial_t^2 Z  \simeq 0  \, .
\ea
Remembering that now the future directed time is $-t$,  the positive frequency mode solution is given by
\ba
Z(r_*, t) = \int_{-\infty}^{+\infty} \frac{\dd k }{\sqrt{2 \pi}} \frac{1}{\sqrt{2 \omega }} \left[
e^{ i ( \omega t  + k r_*) } c_\bfk  +  e^{- i ( \omega t  + k r_*) } c_\bfk^\dagger
\right] \, ,
\ea
in which the new operators $ c_\bfk $ and $c_\bfk ^\dagger$ are defined as the creation and annihilation operators 
for the local observer on left ${\cal I}^-$. 

In terms of the $(u, v)$ coordinate defined in Eq. (\ref{uv-new}),  we have
\ba
\label{wave-rt2}
Z(r_*, t)  = \int_0^\infty \frac{\dd \omega }{\sqrt{2 \pi}} \frac{1}{\sqrt{2 \tilde\omega }} \left[
e^{-i \tilde\omega v }  c_{\tilde \omega}  +  e^{+i \tilde\omega v }  c_{\tilde\omega}^\dagger 
+ e^{-i \tilde\omega u }  c_{-\tilde \omega}  +  e^{+i \tilde\omega u }  c_{-\tilde \omega}^\dagger 
\right] \, .
\ea

Now, concentrating on right movers as before, the outgoing solution near the past horizon $V=0$ and $U>0$ is given by
\ba
\label{Zu-2}
Z(u)  \simeq \int_0^\infty \frac{\dd \tilde\omega }{\sqrt{2 \pi}} \frac{1}{\sqrt{2 \tilde\omega }} \left[
e^{-i \tilde\omega u }  c_{\tilde\omega}  +  e^{+i \tilde\omega u }  c_{\tilde\omega}^\dagger 
\right] \, , \quad \quad ( r \rightarrow 2 GM) \, .
\ea
Note that the plane wave (\ref{Zu-2}) is similar to (\ref{wave-rt}) which were defined for the other half of the past horizon,  
$V=0$ and $U<0$.  The goal is to express $(c_{\tilde \omega}, c_{\tilde\omega}^\dagger)$ in terms of $(a_\omega, a_\omega^\dagger)$ defined for 
a freely falling Kruskal observer equipped with the vacuum (\ref{wave-U}). 

Similar to analysis in subsection \ref{WH-Bogol} we can relate these two sets of operators via the new Bogoliubov coefficients $\overline{\alpha}_{\omega \tilde \omega}$ and $\overline{\beta}_{\omega \tilde \omega}$
as follows
\ba
\label{c-omega2}
c_{\tilde \omega} = \int_0^\infty \dd \omega \Big( \overline{\alpha}_{\omega \tilde \omega}
a_\omega + \overline{\beta}_{\omega \tilde\omega} a_\omega^\dagger
\Big) \, .
\ea

Compared to analysis in subsection  \ref{WH-Bogol}, the difference is that now $U$ changes in the interval $(0, +\infty)$ 
where $U=  e^{\kappa u}/4\kappa$.  The contour in complex plane to calculate the above integral is shown in the right panel of Fig. \ref{contour}.  
Similar to integral (\ref{int-b}) we have 
\ba
\label{int-c}
 \int_{-\infty}^{+\infty} \dd u~ e^{i \tilde \omega u} e^{- i \omega U}  =
 \frac{1}{\kappa} e^{ \frac{\pi \tilde \omega}{2\kappa} } \Big( \frac{\omega}{4\kappa} \Big)^{ \frac{-i\tilde \omega}{\kappa}} \Gamma \big( \frac{i\tilde \omega}{\kappa} \big) \, ,
\ea
and
\ba
 \int_{-\infty}^{+\infty} \dd u~ e^{i \tilde \omega u} e^{ i \omega U}  =  \frac{1}{\kappa} e^{ \frac{-\pi \tilde \omega}{2\kappa} } \Big( \frac{\omega}{4\kappa} \Big)^{ \frac{-i\tilde \omega}{\kappa}} \Gamma \big( \frac{i\tilde \omega}{\kappa} \big) \, ,
\ea
yielding 
\ba
 \left \{ \begin{array}{cc} 
 \overline\alpha_{\omega \tilde\omega} \\ 
 \overline\beta_{\omega \tilde\omega}
\end{array} \right.
= \frac{1}{2 \pi \kappa}  \sqrt{\frac{\tilde \omega}{\omega}}  e^{ \pm \frac{\pi \tilde \omega}{2\kappa} } \,  \Big( \frac{\omega}{4\kappa} \Big)^{ \frac{-i\tilde \omega}{\kappa}} \Gamma \big( \frac{i\tilde \omega}{\kappa} \big) \, .
\ea
Comparing with the values of  $\alpha_{\omega \tilde\omega}$ and $\beta_{\omega \tilde\omega}$ given in  Eq. (\ref{alpha-beta-WH}) we observe 
that
\ba
\overline \alpha_{\omega \tilde\omega} = \alpha_{\omega \tilde\omega}^* \, ,
\quad \quad 
\overline \beta_{\omega \tilde\omega} = \beta_{\omega \tilde\omega}^* \, .
\ea

We see that the results here are on par with the results of the subsections 
\ref{WH-Bogol} and \ref{outside-WH}. In particular, a remote  observer, either on left ${\cal I}^-$ where the perturbations are generated or near the future singularity $r=0$ inside the BH where the perturbations are heading for, conclude that the vacuum defined by a freely falling Kruskal observer (\ref{wave-U})  has 
the particle flux with the spectrum 
\ba
\frac{_{\rm K}\langle 0| c_{ \tilde\omega}^\dagger c_{ \tilde\omega} | 0 \rangle_{K}}{ \delta ({\bf 0})}  = \frac{1}{e^{ \frac{2\pi \tilde \omega}{\kappa} } -1} \, ,
\ea
with the  Hawking temperature $T_H= \frac{\kappa}{2\pi}$ as before.

\section{ Non-vacuum Initial Condition}
\label{C12-general}

In the analysis so far we have assumed that the wave function Eq. (\ref{gk-sol}) inside the WH has 
$C_1=0$ and $C_2=1$. This corresponds to vacuum initial condition. This is similar to Bunch-Davies initial 
condition in inflationary perturbations. Alternatively, we can consider the general case with arbitrary values of
$C_1(\omega)$ and $C_2(\omega)$ subject to the normalization condition Eq. (\ref{norm}). Here we repeat the analysis of Hawking radiation for this general case. We only consider the right mover perturbations generated deep inside the WH region 
propagating towards the past horizon $V=0, U<0$. 

Using the asymptotic form of Hankel function given in Eq. (\ref{Hankel-asym}), 
the mode function near the past horizon inside the WH region is given by
\ba
\label{wave-rt0}
Z(\tau, x)  {|_{r \rightarrow 2 GM }} \rightarrow \int_0^\infty \frac{\dd \omega }{\sqrt{2 \pi}} \frac{1}{\sqrt{2 \omega }} \Big\{ &
 e^{-i \omega v }  \left(C_2 b_\omega + C_1^* b_{-\omega}^\dagger \right)  +   e^{i \omega v }  \left(C_2^* b_\omega^\dagger + C_1 b_{-\omega} \right)  \nonumber\\
&+  e^{-i \omega u }  \left(C_2 b_{-\omega} + C_1^* b_{\omega}^\dagger \right)  +   e^{i \omega u }  \left(C_2^* b_{-\omega}^\dagger + C_1 b_{\omega} \right)  \Big\} \, ,
\ea
in which the coordinates $(u, v)$ are defined as in Eq. (\ref{uv-WH}).  
Also to shorten the notation, we simply wrote $C_i(\omega)= C_i$, but it is understood that the coefficients $C_i$ depend on $\omega$. 

From the above form of mode function we see that one can not uniquely fix the operators $b_{\tilde \omega}$ and $b_{-\tilde\omega}$ just 
by matching the right mover solution $Z(u)$ at the horizon $V=0$. We also need to match the left mover $Z(v)$ at the {\it future} horizon $U=0$ to determine  $b_{\tilde \omega}$ and $b_{-\tilde\omega}$ simultaneously. The left mover mode function near the future horizon  $U=0$ in Kruskal coordinate 
is given by
\ba
\label{wave-V}
Z(V)= \int_0 ^\infty \frac{\dd \omega}{\sqrt{2 \pi} } \frac{1}{\sqrt {2 \omega}} \left( e^{- i \omega V} a_{-\omega} +
e^{ i \omega V} a_{-\omega}^\dagger
\right) \, .
\ea

Performing the inverse Fourier transformation like in Eq. (\ref{b-Omega}) and following the same steps as in Section \ref{WH-Bogol} and imposing the continuity of mode function for both left-movers and right-movers across the past horizon  we obtain 
\ba
\label{b-tilde-non-vac}
b_{\tilde\omega} =   C_2(\tilde \omega)^* Q(\tilde \omega) - C_1(\tilde \omega)^* P(\tilde \omega)^\dagger  ,  \quad \quad 
b_{-\tilde\omega} = C_2(\tilde \omega)^* P(\tilde \omega)  - C_1(\tilde \omega)^* Q(\tilde \omega)^\dagger  ,   \quad 
\ea
in which $P(\tilde \omega)$ and $Q(\tilde \omega)$ are defined via\ba
\label{P-def}
P(\tilde \omega) \equiv \int_0^\infty \dd \omega \Big( \alpha_{\omega \tilde \omega} a_{\omega} + \beta_{\omega \tilde \omega} a_{\omega}^\dagger
\Big) \, ,
\ea
and
\ba
\label{Q-def}
Q(\tilde \omega) \equiv \int_0^\infty \dd \omega \Big( \alpha_{\omega \tilde \omega} a_{-\omega} + \beta_{\omega \tilde \omega} a_{-\omega}^\dagger
\Big) \, ,
\ea
where $( \alpha_{\omega \tilde \omega},  \beta_{\omega \tilde \omega})$ are the Bogoliubov coefficients defined in Section \ref{WH-Bogol}.

Using the normalization condition (\ref{normalization1}) one can check that
\ba
\big[ P(\tilde \omega), P(\tilde \omega')^\dagger]  = \big[ Q(\tilde \omega), Q(\tilde \omega')^\dagger] = \delta (\tilde \omega - \tilde \omega') , \quad
\big[ P(\tilde \omega), Q(\tilde \omega')^\dagger] = \big[ P(\tilde \omega), Q(\tilde \omega')] =
0 \, .
\ea
Note that $b_{\tilde\omega}^\dagger$ and $b_{-\tilde\omega}^\dagger $ are obtained from the  complex conjugation of  $b_{\tilde\omega}$ and $b_{-\tilde\omega}$ in Eq. (\ref{b-tilde-non-vac}). 

Having obtained $b_{-\tilde\omega} $ and $b_{-\tilde\omega}^\dagger $ in terms of $a_{\pm \omega}$ and $a_{\pm \omega}^\dagger$, we are ready to calculate the flux of Hawking radiation as measured by the WH observer or the observer 
outside the WH. From Eq. (\ref{b-tilde-non-vac}) we obtain
\ba
\label{sandwich1}
{_{K}\big\langle 0 \big| b_{ -\tilde\omega}^\dagger b_{ -\tilde\omega} \big| 0 \big\rangle_{K}} =
{_{K}\Big\langle 0 \Big| \Big( C_2(\tilde \omega) P(\tilde \omega)^\dagger - C_1(\tilde \omega) Q(\tilde \omega) \Big) \Big(C_2(\tilde \omega)^* P(\tilde \omega)- C_1(\tilde \omega)^* Q(\tilde \omega)^\dagger \Big) \Big| 0 \Big\rangle_{K}} \, .
\ea
Performing the contractions and using the normalization conditions (\ref{alpha-beta}) and  (\ref{beta-identity}) the flux of particles is obtained to be 
\ba
\label{flux2}
\frac{{_{K}\big\langle 0 \big| b_{ -\tilde\omega}^\dagger b_{ -\tilde\omega} \big| 0 \big\rangle_{K}}}{\delta ({\bf 0})} &=&
\frac{|C_2(\tilde \omega)|^2 +|C_1(\tilde \omega)|^2  e^{\frac{2 \pi \tilde \omega}{\kappa}} }{e^{\frac{2 \pi \tilde \omega}{\kappa}} -1}  \nonumber\\
&=&  \frac{1 }{e^{\frac{2 \pi \tilde \omega}{\kappa}} -1} + |C_1(\tilde \omega)|^2
\coth \left( \frac{\pi \tilde \omega}{\kappa} \right)  \, .
\ea
One can easily see that when $C_1=0$, the above result coincides with the Planck distribution of Hawking radiation. However, when $C_1\neq 0$, we see that the spectrum deviates from the Planck distribution. The deviation from a Planck distribution is significant. It would be interesting to examine if the above result has anything important to say in connection to the information loss problem.

\section{Summary and Discussions}

In  this work we studied the cosmology associated to white hole background. The past singularity inside the WH is like a big bang singularity such that the  WH background represents an anisotropic cosmological setup. The scale factor associated to the extended spatial direction is contracting while the scale factor representing the size of the two-sphere is expanding. Motivated by the cosmological perturbation analysis in FLRW cosmology, in which the observed large scale structures  are created from quantum fluctuations 
during inflation, here we have performed a similar analysis and  studied the quantum perturbations generated deep inside the WH region.  First we considered the case with the vacuum initial condition $C_1=0$. Matching the asymptotic solution to a plane wave solution as measured by a freely falling observer equipped with the Kruskal coordinate and the annihilation and creation operators $a_\bfk$
and $a_\bfk^\dagger$,  we have expressed the annihilation and creation operators $b_\bfk$ and $b_\bfk^\dagger$ defined for an observer deep inside the WH 
in terms of $a_\bfq$ and $a_\bfq^\dagger$.  Furthermore, demanding that the wave function to be continuous across the horizon, we have related   $b_\bfk$
and $b_\bfk^\dagger$ to $d_\bfk$
and $d_\bfk^\dagger$ as defined for an observer outside the WH 
and to ${\cal I}^+$. 

While the observer deep inside the WH and the observer far from BH both agree  on their quantum mode function and share the same vacuum, but their vacuum  is different than the one defined for a freely falling observer near the horizon. In particular, the WH observer and the far observer detect a spectrum of particle with a Planckian distribution and with the Hawking temperature 
$T_H = \frac{1}{8 \pi GM}$. Intuitively speaking, the conclusion is that 
if the BH is not entirely black due to Hawking radiation, then the WH is not entirely white either!

We have extended the analysis to the general case where the initial condition 
deep inside the WH region is not a pure vacuum,  corresponding to 
$C_1 \neq 0$. This may be interpreted as an initial state which contains particles and is not the state of minimum energy. Interestingly, in this case we have shown that the observer far from BH 
detects a flux which deviates from the Planck spectrum with a deviation of the form $|C_1(\tilde \omega)|^2 \coth \left( \frac{\pi \tilde \omega}{\kappa} \right)$.
One has to check if this has interesting implications for information loss problem associated to BH thermodynamics and Hawking radiation. 

To simplify the analysis, we have neglected the scattering of the wave function by the bump of effective potential, practically by setting $\ell=0$. It would be interesting to perform the analysis for the realistic case where the effects of scattering are included. However, we expect that the main results remain unchanged for high frequency modes where the effects of the potential scattering and the dependence of wave function on $\ell$ may not be important. However, for the general case  one expects that the Planck distribution is modified by a greybody  factor. 

\vspace{0.7cm}

{\bf Acknowledgments:}  H. F. would like to thank Misao Sasaki for many insightful discussions and  comments and for collaboration at the early stage of this work. We are grateful to Ted Jacobson, Shinji Mukohyama and Jibril Ben Achour
for useful discussions and comments on the draft.  H.F. would like to thank YITP, Kyoto University for the hospitality during the workshop ``Gravity: Current challenge in black hole physics and cosmology'' where this work was in its final stage. We acknowledge partial support from  the ``Saramadan'' federation of Iran.

\vspace{0.5cm}
  
\appendix 

\section{ Inner product}
\label{inner-prod}

In this appendix we show that the analysis in Section \ref{WH-Bogol} can be obtained using the inner product method as well.

The inner product for the mode functions is defined via \cite{Townsend:1997ku, fabbri2005modeling}
\ba
\label{inner}
(f_\omega,  f_{\omega'}) \equiv  -i \int ( f_\omega f_{\omega', a}^* -  f_{\omega , a} f_{\omega'}^* )~\dd \Sigma^a = \delta (\omega -\omega') 
\ea
in which $d \Sigma^a$ is the initial Cauchy  surface. 

We would like to relate $b_{-\tilde \omega}$ from the mode function of $Z(u)$ in Eq. (\ref{wave-rt})  to $a_\omega$ of the mode function $Z(U)$ in  Eq. (\ref{wave-U}). We choose the surface $d \Sigma^a$ to be the surface near the past event horizon, with $V=0$ and 
$-\infty< U <0$.  

Denoting the positive wavefunction in Eq. (\ref{wave-rt})  as $f_{\tilde \omega} = e^{-i {\tilde \omega} u}/\sqrt{4 \pi {\tilde \omega} } $, we have 
\ba
b_{-{\tilde \omega}} = (Z, f_{{\tilde \omega}}) = - i \int_{-\infty}^0 \dd U \left( Z(U) f_{{\tilde \omega}, U}^* - Z(U)_{, U} f_{\tilde \omega}^*  \right) \, ,
\ea
in which $Z(U)$ is given by Eq. (\ref{wave-U}). 

Now noting that $U= -\frac{1}{4\kappa}e^{-\kappa u}$ we obtain 
\ba
b_{-{\tilde \omega}} =  \int \frac{\dd {\tilde \omega}}{2 \pi} \frac{1}{2 \sqrt{\omega }} \int_{-\infty}^0 \dd U e^{i  u}
 \Big[ (\omega - \frac{4 {\tilde \omega}}{U})  e^{-i \omega U}  a_\omega - 
 (\omega + \frac{4 {\tilde \omega}}{U})  e^{i \omega U} a_\omega^\dagger
\Big] \, .
\ea
Each of the two types of integrals can be calculated, using a similar contour as in Fig.~\ref{contour},  yielding
\ba
\label{int1}
\int_{-\infty}^0 \dd U e^{i {\tilde \omega} u}
  (\omega - \frac{4 {\tilde \omega}}{U})  e^{-i \omega U} = 8 {\tilde \omega} e^{2 \pi {\tilde \omega}} \omega^{4 i {\tilde \omega} } 
   \Gamma(-4 i {\tilde \omega})  \, ,
\ea
and
\ba
\label{int2}
-\int_{-\infty}^0 \dd U e^{i {\tilde \omega} u}
  (\omega + \frac{4 {\tilde \omega}}{U})  e^{i \omega U} = 8 {\tilde \omega} e^{-2 \pi {\tilde \omega}} \omega^{4 i {\tilde \omega} } 
   \Gamma(-4 i {\tilde \omega})   \, .
\ea
Combining these two terms, we obtain $b_{- \tilde \omega}$  in exact agreement with 
Eq. (\ref{b-omega2}) with $\alpha_{\omega \tilde \omega}$ and $\beta_{\omega \tilde \omega}$ given in Eq. (\ref{alpha-beta-WH}).  The rest of analysis is exactly the same as in Section \ref{WH-Bogol}. 


%
\small
\bibliography{references}

\providecommand{\href}[2]{#2}\begingroup\raggedright\begin{thebibliography}{10}

\bibitem{LIGOScientific:2016aoc}
{\scshape LIGO Scientific, Virgo} collaboration, B.~P. Abbott et~al.,
  \emph{{Observation of Gravitational Waves from a Binary Black Hole Merger}},
  \href{https://doi.org/10.1103/PhysRevLett.116.061102}{\emph{Phys. Rev. Lett.}
  {\bfseries 116} (2016) 061102},
  [\href{https://arxiv.org/abs/1602.03837}{{\ttfamily 1602.03837}}].

\bibitem{LIGOScientific:2016sjg}
{\scshape LIGO Scientific, Virgo} collaboration, B.~P. Abbott et~al.,
  \emph{{GW151226: Observation of Gravitational Waves from a 22-Solar-Mass
  Binary Black Hole Coalescence}},
  \href{https://doi.org/10.1103/PhysRevLett.116.241103}{\emph{Phys. Rev. Lett.}
  {\bfseries 116} (2016) 241103},
  [\href{https://arxiv.org/abs/1606.04855}{{\ttfamily 1606.04855}}].

\bibitem{LIGOScientific:2018jsj}
{\scshape LIGO Scientific, Virgo} collaboration, B.~P. Abbott et~al.,
  \emph{{Binary Black Hole Population Properties Inferred from the First and
  Second Observing Runs of Advanced LIGO and Advanced Virgo}},
  \href{https://doi.org/10.3847/2041-8213/ab3800}{\emph{Astrophys. J. Lett.}
  {\bfseries 882} (2019) L24},
  [\href{https://arxiv.org/abs/1811.12940}{{\ttfamily 1811.12940}}].

\bibitem{Sakharov:1966aja}
A.~D. Sakharov, \emph{{Nachal'naia stadija rasshirenija Vselennoj i
  vozniknovenije neodnorodnosti raspredelenija veshchestva}}, {\emph{Sov. Phys.
  JETP} {\bfseries 22} (1966) 241}.

\bibitem{Poisson:1988wc}
E.~Poisson and W.~Israel, \emph{{Structure of the Black Hole Nucleus}},
  \href{https://doi.org/10.1088/0264-9381/5/12/002}{\emph{Class. Quant. Grav.}
  {\bfseries 5} (1988) L201--L205}.

\bibitem{Frolov:1989pf}
V.~P. Frolov, M.~A. Markov and V.~F. Mukhanov, \emph{{Through a Black Hole into
  a New Universe?}},
  \href{https://doi.org/10.1016/0370-2693(89)91114-3}{\emph{Phys. Lett. B}
  {\bfseries 216} (1989) 272--276}.

\bibitem{Frolov:1988vj}
V.~P. Frolov, M.~A. Markov and V.~F. Mukhanov, \emph{{Black Holes as Possible
  Sources of Closed and Semiclosed Worlds}},
  \href{https://doi.org/10.1103/PhysRevD.41.383}{\emph{Phys. Rev. D} {\bfseries
  41} (1990) 383}.

\bibitem{Firouzjahi:2016nle}
H.~Firouzjahi, \emph{{Primordial Universe Inside the Black Hole and
  Inflation}},  \href{https://arxiv.org/abs/1610.03767}{{\ttfamily
  1610.03767}}.

\bibitem{wenda1988junction}
S.~Wenda and S.~Zhu, \emph{Junction conditions on null hypersurface},
  {\emph{Physics Letters A} {\bfseries 126} (1988) 229--232}.

\bibitem{Brandenberger:2021ken}
R.~Brandenberger, L.~Heisenberg and J.~Robnik, \emph{{Non-singular black holes
  with a zero-shear S-brane}},
  \href{https://doi.org/10.1007/JHEP05(2021)090}{\emph{JHEP} {\bfseries 05}
  (2021) 090}, [\href{https://arxiv.org/abs/2103.02842}{{\ttfamily
  2103.02842}}].

\bibitem{Gaztanaga:2021rnv}
E.~Gaztanaga, \emph{{Inside a Black Hole: the illusion of a Big Bang}},
  \href{https://hal.archives-ouvertes.fr/hal-03106344v7}{\emph{Hal Open
  Science} {\bfseries 7} (2021) 03783110}.

\bibitem{Gaztanaga:2022gbd}
E.~Gaztanaga, \emph{{How the Big Bang Ends Up Inside a Black Hole}},
  \href{https://doi.org/10.3390/universe8050257}{\emph{Universe} {\bfseries 8}
  (2022) 257}, [\href{https://arxiv.org/abs/2204.11608}{{\ttfamily
  2204.11608}}].

\bibitem{Sato:1981bf}
K.~Sato, M.~Sasaki, H.~Kodama and K.-i. Maeda, \emph{{Creation of Wormholes by
  First Order Phase Transition of a Vacuum in the Early Universe}},
  \href{https://doi.org/10.1143/PTP.65.1443}{\emph{Prog. Theor. Phys.}
  {\bfseries 65} (1981) 1443}.

\bibitem{Maeda:1981gw}
K.-i. Maeda, K.~Sato, M.~Sasaki and H.~Kodama, \emph{{Creation of De
  Sitter-schwarzschild Wormholes by a Cosmological First Order Phase
  Transition}}, \href{https://doi.org/10.1016/0370-2693(82)91151-0}{\emph{Phys.
  Lett. B} {\bfseries 108} (1982) 98--102}.

\bibitem{Sato:1981gv}
K.~Sato, H.~Kodama, M.~Sasaki and K.-i. Maeda, \emph{{Multiproduction of
  Universes by First Order Phase Transition of a Vacuum}},
  \href{https://doi.org/10.1016/0370-2693(82)91152-2}{\emph{Phys. Lett. B}
  {\bfseries 108} (1982) 103--107}.

\bibitem{Farhi:1986ty}
E.~Farhi and A.~H. Guth, \emph{{An Obstacle to Creating a Universe in the
  Laboratory}}, \href{https://doi.org/10.1016/0370-2693(87)90429-1}{\emph{Phys.
  Lett. B} {\bfseries 183} (1987) 149--155}.

\bibitem{Blau:1986cw}
S.~K. Blau, E.~I. Guendelman and A.~H. Guth, \emph{{The Dynamics of False
  Vacuum Bubbles}}, \href{https://doi.org/10.1103/PhysRevD.35.1747}{\emph{Phys.
  Rev. D} {\bfseries 35} (1987) 1747}.

\bibitem{Oshita:2016btk}
N.~Oshita and J.~Yokoyama, \emph{{Creation of an inflationary universe out of a
  black hole}},
  \href{https://doi.org/10.1016/j.physletb.2018.08.018}{\emph{Phys. Lett. B}
  {\bfseries 785} (2018) 197--200},
  [\href{https://arxiv.org/abs/1601.03929}{{\ttfamily 1601.03929}}].

\bibitem{markov1982limiting}
M.~Markov, \emph{Limiting density of matter as a universal law of nature},
  {\emph{JETP Lett.(Engl. Transl.);(United States)} {\bfseries 36} (1982) }.

\bibitem{Kantowski:1966te}
R.~Kantowski and R.~K. Sachs, \emph{{Some spatially homogeneous anisotropic
  relativistic cosmological models}},
  \href{https://doi.org/10.1063/1.1704952}{\emph{J. Math. Phys.} {\bfseries 7}
  (1966) 443}.

\bibitem{Chakraborty:2015nwa}
S.~Chakraborty, S.~Singh and T.~Padmanabhan, \emph{{A quantum peek inside the
  black hole event horizon}},
  \href{https://doi.org/10.1007/JHEP06(2015)192}{\emph{JHEP} {\bfseries 06}
  (2015) 192}, [\href{https://arxiv.org/abs/1503.01774}{{\ttfamily
  1503.01774}}].

\bibitem{Eardley:1974zz}
D.~M. Eardley, \emph{{Death of White Holes in the Early Universe}},
  \href{https://doi.org/10.1103/PhysRevLett.33.442}{\emph{Phys. Rev. Lett.}
  {\bfseries 33} (1974) 442--444}.

\bibitem{Jusufi:2017nib}
K.~Jusufi, \emph{{Hawking Radiation in the Spacetime of White Holes}},
  \href{https://doi.org/10.1007/s10714-018-2406-0}{\emph{Gen. Rel. Grav.}
  {\bfseries 50} (2018) 84},
  [\href{https://arxiv.org/abs/1711.09625}{{\ttfamily 1711.09625}}].

\bibitem{Giavoni:2020gui}
C.~Giavoni and M.~Schneider, \emph{{Quantum effects across dynamical
  horizons}}, \href{https://doi.org/10.1088/1361-6382/abb576}{\emph{Class.
  Quant. Grav.} {\bfseries 37} (2020) 215020},
  [\href{https://arxiv.org/abs/2003.11095}{{\ttfamily 2003.11095}}].

\bibitem{Barcelo:2016hgb}
C.~Barcel\'o, R.~Carballo-Rubio and L.~J. Garay, \emph{{Exponential fading to
  white of black holes in quantum gravity}},
  \href{https://doi.org/10.1088/1361-6382/aa6962}{\emph{Class. Quant. Grav.}
  {\bfseries 34} (2017) 105007},
  [\href{https://arxiv.org/abs/1607.03480}{{\ttfamily 1607.03480}}].

\bibitem{Regge:1957td}
T.~Regge and J.~A. Wheeler, \emph{{Stability of a Schwarzschild singularity}},
  \href{https://doi.org/10.1103/PhysRev.108.1063}{\emph{Phys. Rev.} {\bfseries
  108} (1957) 1063--1069}.

\bibitem{Hawking:1975vcx}
S.~W. Hawking, \emph{{Particle Creation by Black Holes}},
  \href{https://doi.org/10.1007/BF02345020}{\emph{Commun. Math. Phys.}
  {\bfseries 43} (1975) 199--220}.

\bibitem{Unruh:1976db}
W.~G. Unruh, \emph{{Notes on black hole evaporation}},
  \href{https://doi.org/10.1103/PhysRevD.14.870}{\emph{Phys. Rev. D} {\bfseries
  14} (1976) 870}.

\bibitem{Unruh:1983ms}
W.~G. Unruh and R.~M. Wald, \emph{{What happens when an accelerating observer
  detects a Rindler particle}},
  \href{https://doi.org/10.1103/PhysRevD.29.1047}{\emph{Phys. Rev. D}
  {\bfseries 29} (1984) 1047--1056}.

\bibitem{Mukhanov:2007zz}
V.~Mukhanov and S.~Winitzki, \emph{{Introduction to quantum effects in
  gravity}}.
\newblock Cambridge University Press, 6, 2007.

\bibitem{Townsend:1997ku}
P.~K. Townsend, \emph{{Black holes: Lecture notes}},
  \href{https://arxiv.org/abs/gr-qc/9707012}{{\ttfamily gr-qc/9707012}}.

\bibitem{fabbri2005modeling}
A.~Fabbri and J.~Navarro-Salas, \emph{Modeling black hole evaporation}.
\newblock World Scientific, 2005.

\bibitem{Visser:2001kq}
M.~Visser, \emph{{Essential and inessential features of Hawking radiation}},
  \href{https://doi.org/10.1142/S0218271803003190}{\emph{Int. J. Mod. Phys. D}
  {\bfseries 12} (2003) 649--661},
  [\href{https://arxiv.org/abs/hep-th/0106111}{{\ttfamily hep-th/0106111}}].

\bibitem{Jacobson:2003vx}
T.~Jacobson, \emph{{Introduction to quantum fields in curved space-time and the
  Hawking effect}},  in \emph{{School on Quantum Gravity}}, pp.~39--89, 8,
  2003, \href{https://arxiv.org/abs/gr-qc/0308048}{{\ttfamily gr-qc/0308048}},
  \href{https://doi.org/10.1007/0-387-24992-3_2}{DOI}.

\end{thebibliography}\endgroup
\bibliographystyle{JHEP}

\end{document}